\newcommand{\PaperTitle}{Lazy Eye Inspection: Capturing the State of Happy Eyeballs Implementations}

\documentclass[sigconf]{acmart}

\copyrightyear{2025}
\acmYear{2025}
\setcopyright{cc}
\setcctype{by}
\acmConference[IMC '25]{Proceedings of the 2025 ACM Internet Measurement Conference}{October 28--31, 2025}{Madison, WI, USA}
\acmBooktitle{Proceedings of the 2025 ACM Internet Measurement Conference (IMC '25), October 28--31, 2025, Madison, WI, USA}
\acmDOI{10.1145/3730567.3732925}
\acmISBN{979-8-4007-1860-1/2025/10}

\usepackage{packages}
\usepackage{wasysym}
\usepackage{threeparttable}

\makeatletter
\newcommand\EatSpacesHack{\@bsphack\@esphack}
\makeatother
\iffalse{}
\newcommand\reviewfix[1]{{\color{red}\sffamily\bfseries [RF:\ #1]}\EatSpacesHack}
\else
\newcommand\reviewfix[1]{\EatSpacesHack}
\fi
\makeatother

\newcommand{\jz}[1]{\hl{\textbf{JZ: #1}}}
\newcommand{\ps}[1]{\hl{\textbf{PS: #1}}}
\newcommand{\lw}[1]{\hl{\textbf{LW: #1}}}
\newcommand{\mk}[1]{\hl{\textbf{MK: #1}}}

\renewcommand{\jz}[1]{\iffalse\hl{\textbf{JZ: #1}}\fi}
\renewcommand{\ps}[1]{\iffalse\hl{\textbf{PS: #1}}\fi}
\renewcommand{\lw}[1]{\iffalse\hl{\textbf{LW: #1}}\fi}
\renewcommand{\mk}[1]{\iffalse\hl{\textbf{MK: #1}}\fi}

\newcommand{\hevone}{\ac{he}v1\xspace}

\newcommand{\hevthree}{\ac{he}v3\xspace}
\newcommand{\cad}{\ac{cad}\xspace}
\newcommand{\rd}{\ac{rd}\xspace}
\newcommand{\aaaar}{\texttt{AAAA}\xspace}
\newcommand{\ar}{\texttt{A}\xspace}
\newcommand{\svcbr}{\texttt{SVCB}\xspace}
\newcommand{\httpsr}{\texttt{HTTPS}\xspace}
\newcommand{\tabbullet}{{\large$\bullet$}}
\newcommand{\tabcirc}{{\large$\circ$}}
\newcommand{\tableftcirc}{{\large$\leftcirc$}}
\newcommand{\tabrightcirc}{{\large$\rightcirc$}}
\newcommand{\tabdowncirc}{{\large$\downcirc$}}

\def\leftcirc{\raisebox{1pt}{\scalebox{0.6}{\LEFTcircle}}}
\def\rightcirc{\raisebox{1pt}{\scalebox{0.6}{\RIGHTcircle}}}
\def\downcirc{\raisebox{1pt}{\scalebox{0.6}{\rotatebox{90}{\RIGHTcircle}}}}

\begin{document}

\title{\PaperTitle}

\author{Patrick Sattler}
\email{sattler@net.in.tum.de}
\orcid{1234-5678-9012}
\affiliation{%
  \institution{Technical University of Munich}
  \city{Munich}
  \country{Germany}
}

\author{Matthias Kirstein}
\email{kirstein@net.in.tum.de}
\orcid{0009-0008-5122-1950}
\affiliation{%
  \institution{Technical University of Munich}
  \city{Munich}
  \country{Germany}
}

\author{Lars Wüstrich}
\email{wuestrich@net.in.tum.de}
\orcid{0009-0004-6462-4127}
\affiliation{%
  \institution{Technical University of Munich}
  \city{Munich}
  \country{Germany}
}

\author{Johannes Zirngibl}
\email{jzirngib@mpi-inf.mpg.de}
\orcid{0000-0002-2918-016X}
\affiliation{%
  \institution{Max Planck Institute for Informatics}
  \city{Saarbrücken}
  \country{Germany}
}

\author{Georg Carle}
\email{carle@net.in.tum.de}
\orcid{0000-0002-2347-1839}
\affiliation{%
  \institution{Technical University of Munich}
  \city{Munich}
  \country{Germany}
}

\begin{abstract}
\reviewfix{4-1} While transitioning to an IPv6-only communication, many devices settled on a dual-stack setup.
IPv4 and IPv6 are available to these hosts for new connections.
Happy Eyeballs (HE) describes a mechanism to prefer IPv6 for such hosts while ensuring a fast fallback to IPv4 when IPv6 fails.
The IETF is currently working on the third version of HE.
While the standards include recommendations for HE parameter choices, it is up to the client and OS to implement HE.
In this paper, we investigate the state of HE in various clients, particularly web browsers and recursive resolvers.
We introduce a framework to analyze and measure clients' HE implementations and parameter choices.
According to our evaluation, only Safari supports all HE features.
Safari is also the only client implementation in our study that uses a dynamic IPv4 connection attempt delay, a resolution delay, and interlaces addresses.
We further show that problems with the DNS A record lookup can even delay and interrupt the network connectivity despite a fully functional IPv6 setup with Chrome and Firefox.
We operate a publicly available website (\url{www.happy-eyeballs.net}) which measures the browser's HE behavior, and we publish our testbed measurement framework.
\end{abstract}



\maketitle

\section{Introduction}
\label{sec:introduction}

The standardization of IPv6~\cite{rfc2460} as a non-compatible protocol to IPv4 resulted in a separation of the IP layer.
Due to the IPv4 address space exhaustion, IPv6 needs to be deployed and used more widely.
While the general development is towards dual-stack devices, the non-compatibility imposes a decision on clients: Which address family should be used?
However, the Internet has yet to fully transform~\cite{apnicipv6,googlev6growth, Streibelt2023IPv6DNS}, thus, the decision is nontrivial. %
To improve IPv6 usage without drastically influencing user experience, \ac{he} was designed~\cite{rfc6555}.
Across two versions~\cite{rfc6555,rfc8305}, \ac{he} provides suggestions for different parts of connections on how to prefer IPv6 but fall back on IPv4 if necessary.

Since IPv6, new transport layer protocols have been standardized, resulting in further separations on additional ISO/OSI stack layers.
QUIC~\cite{rfc9000} resulted in another transport protocol besides TCP and the foundation for a new version of HTTP.
Like IPv6, QUIC and HTTP/3 are incompatible with existing alternatives, and clients must decide which protocol to use.
Therefore, as of November 2024, a new version of \ac{he}~\cite{draft-pauly-v6ops-happy-eyeballs-v3} and a new IETF working group~\cite{happy-charter} are proposed to provide guidance given the new complexity.

However, the existing \ac{he} ecosystem is not yet understood.
All major browsers have implemented \ac{he}~\cite{Huston2012HE} in theory. %
Nevertheless, both RFCs only provide suggestions, and how browsers and other clients, \eg{} resolvers, implement \ac{he} has yet to be documented.
With the increasing complexity of the Internet and the attempt to standardize a new, more complex version of \ac{he}, it is essential to understand the current RFC implementations.
Clients' decisions highly impact the user experience, the visibility of network issues, and potential deployment errors.

\noindent
Our main contributions in this paper are:

\first{} An evaluation of \ac{he} implementations in nine browsers on seven \acp{os}.
We show that current implementations vary widely and are mostly limited to the first version of \ac{he} (RFC6555~\cite{rfc6555}).

\second{} \reviewfix{4-2} A measurement of resolver IP selection and fallback behavior. We observe that this critical network component generally does not rely on \ac{he} style approaches in a dual-stack setup. Our results show a wide variety of behavior with almost no operators using a strict IPv6 version preference.

\third{} \reviewfix{4-2} A local test framework that tests the \ac{he} implementations of multiple clients across different versions.

\fourth{} A web-based testing tool that emulates network delays for all protocols \ac{he} considers.
It is helpful to test browsers and resolvers on different \acp{os}.
Moreover, anyone can use our public tool to test the browser of their choice.

The website to access the tool, the source code for the test framework, and the web-based tool are available at:

\begin{center}
    \textbf{\url{https://www.happy-eyeballs.net/}}
\end{center}

\begin{figure}[b]
    \includegraphics{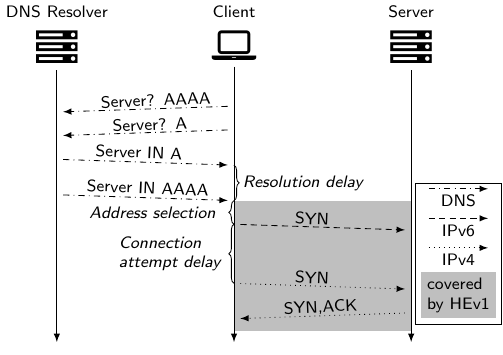}
    \caption{Happy Eyeballs (v2): A client first resolves the server's name.
    Then it prioritizes the addresses based on a local policy and proceeds to attempt to establish a connection to the server.
    HEv1 only focuses on the connection establishment, HEv2 additionally includes the name resolution.}
    \label{fig:he}
\end{figure}

\begin{table*}
    \small
    \caption{Comparison of parameters defined for \ac{he}v1~\cite{rfc6555}, v2~\cite{rfc8305} and the draft for v3~\cite{draft-pauly-v6ops-happy-eyeballs-v3} as of April 2025. \reviewfix{5}}
    \label{tab:he_parameters}
    \begin{tabular}{lccccc}
        \toprule
        Parameter & HEv1 (2012) & HEv2 (2017) & HEv3 (2025-ongoing) & Local tests & Web tests \\
        \midrule
        Considered protocols & IPv4, IPv6 & IPv4, IPv6, DNS & IPv4, IPv6, DNS, QUIC & $\checkmark$ & $\checkmark$\\
        \midrule
        DNS Records& - & \aaaar{}, \ar{} & \svcbr{}, \httpsr{}, \aaaar{}, \ar{} & $\checkmark$ & $\checkmark$ \\
        Resolution Delay & - & \SI{50}{\milli\second} & \SI{50}{\milli\second} & $\checkmark$ & $\checkmark$ \\
        \midrule
        Address selection & IPv6 once, then IPv4 & alternating & IP family and L4 protocol & $\checkmark$  & - \\
        \midrule
        Fixed Conn. Attempt Delay  & \si{150}-\SI{250}{\milli\second} & \SI{250}{\milli\second} & \SI{250}{\milli\second} & $\checkmark$ & $\checkmark$\\
        $\mapsto$Min/Rec./Max when dynamic  & - & \SI{10}{\milli\second} / \SI{100}{\milli\second} / \SI{2}{\second}& \SI{10}{\milli\second} / \SI{100}{\milli\second} / \SI{2}{\second} & $\checkmark$ & $\checkmark$\\
        \bottomrule
    \end{tabular}
\end{table*}

\section{Happy Eyeballs}
\label{sec:he-background}

\acf{he} aims to enhance the user experience by allowing a fast fallback to IPv4 in cases of impaired IPv6 connectivity.
The IETF has standardized two \ac{he} versions~\cite{rfc6555, rfc8305}, and version 3~\cite{draft-pauly-v6ops-happy-eyeballs-v3} is in draft.
This section introduces their algorithms (see \Cref{fig:he}) and explains their configuration parameters (see \Cref{tab:he_parameters}). %

\textbf{HEv1}
(2012)~\cite{rfc6555}, focuses on basic connection attempts from a client to a server~(gray area in \Cref{fig:he}).
The idea is to race the connection establishment between IPv6 and IPv4 addresses while prioritizing one protocol family based on the host's address selection policy.
If DNS returns multiple records for each address family, the host orders them according to its address selection policy.
\ac{he}v1 recommends falling back to IPv4 after the first IPv6 attempt, even if multiple IPv6 addresses are available.
If the initial IPv6 and IPv4 connections fail, \ac{he}v1 does not provide further instructions on the order of subsequent connection attempts, leaving this decision up to the application developers.

The core parameter of \ac{he}v1 is the \emph{\cad{}}, which is recommended to be within \si{150}-\SI{250}{\milli\second}~(see \Cref{tab:he_parameters}).
If an application's IPv6 connection is not successfully established within this delay, the application should initiate a parallel IPv4 connection.
\ac{he} largely avoids harming the network or servers, as the second connection is only initiated if IPv6 appears slow or broken.
Once one connection attempt succeeds, the client discards the others and should cache the outcome for ``the order of \SI{10}{\minute}''~\cite{rfc6555}.

\textbf{HEv2}
 (2017)~\cite{rfc8305} additionally considers DNS~(the complete process in Figure~\ref{fig:he}) and extends the description of the \emph{address selection}.
In \ac{he}v2, a client should first issue a \aaaar{} query, immediately followed by an A query.
If the client receives a response with \aaaar{} records first, it immediately starts a connection attempt to the target host using IPv6.
If an A response arrives first, the application should wait for a \emph{\rd{}} of \SI{50}{\milli\second} that starts with the reception of the A record (see \Cref{tab:he_parameters}).
If the client receives a \aaaar{} response during the \rd{}, a connection attempt via IPv6 is immediately initiated.
The client starts a connection attempt via IPv4 after the \rd if there is no \aaaar{} answer.
This behavior prioritizes IPv6 while minimizing the user-visible delay if it is impaired.

The next step in \ac{he}v2 is the address selection of the retrieved addresses by using a predefined policy.
In addition to IP address family, clients can leverage knowledge about historical TCP round-trip times and previously used addresses.
Finally, the addresses in the list are interlaced to alternate between IP address families.
The first address at the top of the list should be an IPv6 address to maintain the IPv6 preference.
The list may start with multiple IPv6 addresses. %
The number of those addresses is called the \emph{First Address Family Count}, which is recommended to be $1$ or $2$ for aggressively favoring one family~(see \Cref{tab:he_parameters}).
Like in \ac{he}v1, the client then proceeds to establish a connection using the previously selected addresses.
It uses the \cad{} between subsequent attempts to prevent network congestion.
\ac{he}v2 recommends \SI{250}{\milli\second} for a fixed \cad{}.
For history-inferred implementations, it defines an absolute minimum \cad{} of \SI{10}{\milli\second}, recommends a minimum of \SI{100}{\milli\second}, and a maximum \cad{} of \SI{2}{\second}~(see \Cref{tab:he_parameters}).

\textbf{HEv3}
As of 2025, the IETF is working on \ac{he}v3~\cite{draft-pauly-v6ops-happy-eyeballs-v3}.
\ac{he}v3 will process existing \svcbr{} or \httpsr{} DNS resource records~\cite{rfc9460,Zirngibl2023HTTPS} to support protocol discovery.
The \ac{he}v3 address selection should favor IP addresses with available TLS Encrypted ClientHello (ECH) over QUIC over TCP.
Besides those changes, \ac{he}v3 is currently similar to \ac{he}v2, using the same parameters and recommendations.

\section{Related Work}
\label{sec:related-work}

To the best of our knowledge, the actual implementation of \ac{he} in clients has not been researched, except for initial articles during the standardization of \ac{he}v1, \eg{} by \citeauthor{Aben2011HE}~\cite{Aben2011HE} or \citeauthor{Huston2012HE}~\cite{Huston2012HE}.
The effects of \ac{he} have often been analyzed based on the performance of IPv6 and IPv4 deployments on the Internet and the suggestions from the \ac{he} RFCs~\cite{rfc6555,rfc8305} (see \Cref{tab:he_parameters}).
\citeauthor{bajpai2016he}~\cite{bajpai2016he,bajpai2019he} evaluated the latency towards services based on the protocol used and inferred which protocol would be used by clients based on the delays from RFCs.
\citeauthor{Pujol2017Capabilities}~\cite{Pujol2017Capabilities} evaluated client and service IPv6 capabilities.
Based on network metrics, they argued that Happy Eyeballs is visible, but most clients should rely on IPv6 based on suggested delays.
However, our work shows that the suggestions from RFCs are rarely used, and \ac{he} is only partially implemented.
Different studies~\cite{apnicipv6,googlev6growth,Streibelt2023IPv6DNS} have shown a trend towards an IPv6-only Internet, but essential components still lack full IPv6 support.
Therefore, \ac{he}v1 and v2, focusing on IPv6, are still important mechanisms for the quality of experience on the Internet. %

While IPv6 deployment started slowly, \citeauthor{Zirngibl2021Over9000}~\cite{Zirngibl2021Over9000, Zirngibl2024QUIC} have shown early and widespread QUIC deployment.
\citeauthor{Dong2024HTTPS}~\cite{Dong2024HTTPS} have shown that \httpsr{} DNS resource records, including the first ECH information, are available.
Therefore, an early standardization of \ac{he}v3 considering these possibilities is reasonable.
However, \citeauthor{Dong2024HTTPS}~\cite{Dong2024HTTPS} show that most browsers do not fully use information from \httpsr{} resource records as of 2024.
Our work shows that most browsers currently do not implement \ac{he}v2 yet.
\reviewfix{3} \citeauthor{foremski2019observatory}~\cite{foremski2019observatory} analyze the effect of small negative caching values on clients using \ac{he} and find domains with up to \sperc{90} empty \texttt{AAAA} responses due to \ac{he}.
Similar to \ac{he}v3 drafts, \citeauthor{Papastergiou2016TransportHE}~\cite{Papastergiou2016TransportHE} suggested racing different transport protocols alongside the IP layer.
They implement an example to show potential advantages, but do not consider current \ac{he} implementations and whether they can be combined.
Our findings should be considered during the standardization and  evaluation of the Internet and the potential effects of \ac{he} in the future.
Especially if the overall algorithm gets more complex, clients might drift farther from suggestions.

\section{Measurement Methodology}
\label{sec:approach}

We set up two test environments to get insight into how different clients use \ac{he}: a local, fully controlled one and a web-based tool.
Our approach treats the software under test as a black box.
Therefore, it is usable with any software that can target arbitrary targets (\eg command line tools).

\subsection{Measurement Targets}
First, we define the measurement targets and our approach to evaluate these.
\ac{he} defines \emph{Configurable Values}.
We evaluate \ac{he} clients along these values.
\Cref{tab:he_parameters} shows these variables and the environment in which we measured them. %
The following describes how we target the \cad{}, the \rd{}, the address selection approach, and the resolver behavior.
\reviewfix{1.1} The observation method depends on the measurement setup and is described in \Cref{ssec:measurement-setup}.

\textit{\first{} Connection Attempt Delay:}
\reviewfix{1}
The \cad{} is the time the client waits for an IPv6 response before initiating a parallel connection attempt via IPv4.
To obtain the \cad{}, we configure a domain name to resolve to an IPv4 and IPv6 address.
We simulate a slower IPv6 connection by delaying IPv6 packets using \texttt{tc-netem} \cite{netem} on the server side.
If the introduced delay exceeds the client's \cad, it should initiate an IPv4 connection.
We can observe these attempts and infer the \cad.

\textit{\second{} Resolution Delay:}
The \rd{} is the delay the client waits for an \aaaar (IPv6 address record) query response after it obtained the \ar (IPv4 address record) query response.
We use a custom authoritative name server implementation to delay responses according to encoded test parameters.
These parameters include the delay, the resource record type to delay, and a nonce to prevent caching effects. %
This approach allows a variety of different test setups with a single server deployment.

\textit{\third{} Address Selection:}
Our DNS-based measurements show whether both record types are queried and if the client has any \ac{he} \cad configured.
We extend this test case to include multiple addresses per IP family to evaluate the destination address selection by clients.
We use addresses that do not respond at all, and thus should trigger the fallback mechanism to other addresses.

\subsection{Recursive Resolver:}
\reviewfix{2}
The \ac{he} standardization primarily targets end-user client software, and as part of these mainly browser maintainers are involved.
While the specifics of the \ac{he} standardization (\eg recommended delay values) are not generalizable, we can still apply our measurement setup to other protocols and applications to learn about their IP version selection and fallback behavior.
The results can then be interpreted in the context of the analyzed use case.
Our measurement approach can be expanded to include transport protocols in the future.
The DELEG working group~\cite{deleg} discusses the indication of these protocols to resolvers.

In this work, we will look at the behavior of DNS resolver software.
Recursive resolvers play a central role in Internet communication.
They must also decide which IP version to use during the iterative resolution process.
At each delegation to a new zone, and thus usually a new authoritative name server, the recursive resolver decides if and how to prefer IPv6 over IPv4 to connect to the zone's name server.

Therefore, we extended our measurement setup and applied it to the resolver software to check their IP preference behavior in general and if they use \ac{he} style approaches.
We need to apply our traffic shaping to the authoritative name servers serving our zones to enable such measurements.
Instead of different domain names inside a single zone, we created entirely different zones for each measured delay.
Our traffic shaping is applied to the name server records (in case of \rd measurements) and the corresponding IP addresses (for \cad measurements).
Recursive resolvers often include performance metrics when deciding which name server and IP address to use.
Our authoritative IP addresses are not used by any zone outside our measurement campaign.
Additionally, we use unique zone apexes and unique authoritative name server names to reduce the impact of such caching effects.

\subsection{Measurement Setups}
\label{ssec:measurement-setup}
We implement these test cases in two different setups:

\textit{\first{} Local Testbed Framework:}
We set up  a local testbed of two directly connected hosts (client and server) to eliminate any network effects in our measurements.
This setup is an ideal scenario where IPv4 and IPv6 connections between two dual-stacked hosts experience equal delays.
We use a separate interface to set up and send instructions to the hosts.

A controller process runs our framework process to initiate measurement runs.
Our framework is extendable, as the test cases and clients to test are defined outside the framework's code (see App.~\Cref{fig:local_testbed}).
They consist of a config file instructing the framework which executables to run when.
For example, to set up our \cad test, we start an NGINX instance~\cite{nginx} and an authoritative DNS server and configure delays using \texttt{tc-netem} on the server host.
On the client host, we start a packet capture, the client application, and initiate the request.
Setup stages can be executed at each test run configuration (\eg delay), or only at the start and end of a test case.
The test configuration also supports defining the ranges and intervals for the test run configuration variable to execute coarse initial runs and fine-grained follow-ups.
The clients are defined separately from the test cases and start via predefined script calls, which each client must implement independently.
Our setup uses containers to start with a clean client state and use Selenium~\cite{selenium} to control browsers.
\reviewfix{3} After each test case configuration, we reset the client to a predefined state (\eg drop and create a new container) to prevent any caching effects.

We determine the \cad by measuring the time between the first IPv6 packet and the first IPv4 packet observed in the client's packet capture.
The local testbed's accuracy depends on the packet capture time stamping accuracy, which is usually below a millisecond~\cite{gallenmueller2020pcapaccuracy}.
Therefore, this method gives us the needed accuracy for the \cad value as real-world connections usually expose larger jitter than our measurement accuracy.

\textit{\second{} Web-based Testing Tool:}
We created a web-based tool to enable tests from a broader set of browsers in combination with different operating systems.
This tool also allows testing on devices that cannot be directly attached to a server and remotely instructed to perform specific tasks.
The results can be used to validate our local testbed measurements regarding real-world network conditions. %

We evaluate results solely from the client side.
\reviewfix{1} Our web server returns the client's source address in its response, and the client uses the response to determine the used IP version.
The nature of this web deployment does not allow resetting client and server configurations after each measurement.
Therefore, we use a fixed set of 18 delays between 0 and \SI{5}{\second} (see App. \Cref{fig:cadweb_screenshot}).
Each delay has dedicated IPv4 and IPv6 addresses assigned, and we delay IPv6 traffic accordingly.
Therefore, the \cad can only be determined to be in the interval of the last delay using IPv6 and the first delay using IPv4, e.g., the \cad{} for Safari in App. \Cref{fig:cadweb_screenshot} is $CAD \in \mathopen(200,250\mathclose]$.
Nevertheless, it is valuable to obtain results in actual network conditions.
Some clients adjust their behavior in non-lab conditions, significantly influencing observed results (see \Cref{sec:cad}).
Furthermore, we associate a dedicated domain to each delay-address pair to prevent caching.
We reuse our custom DNS setup with a publicly resolvable zone.

%

\section{Evaluation}
\label{sec:eval}

We evaluate our measurement targets based on results obtained via the local testbed and the web-based testing tool.

\subsection{Connection Attempt Delay}
\label{sec:cad}

Our local testbed includes Mozilla Firefox, Google Chrome, Microsoft Edge, and Chromium.
All browsers were tested on the latest Ubuntu LTS image (\textit{24.04.1}).
We included browser versions as old as January 2021.
Additionally, we measured Safari (17.5) on \textit{MacOS}.
This is a set of the most used browsers~\cite{browsermarketshare,w3browserstatistics} and, thus, a solid  base for our web-based evaluation.
Moreover, most alternative browsers are Chromium-based, and they behave the same.
In addition to browsers, we included tests for curl~\cite{curl} and wget~\cite{wget} as popular command-line tools.

\reviewfix{1.1} Initial measurements using coarse delay configurations indicate the rough time interval of the configured \cad{}.
Except for Safari, all clients stayed within \SI{300}{\milli\second}.
Our fine-grained delay configurations perform measurements from 0 to \SI{400}{\milli\second} in \SI{5}{\milli\second} steps.
It is important to note that any local measurement that uses a delay larger than the client's \cad also observes the \cad value in the packet capture.
Our coarse evaluation measures the \cad at least 20 times per client and testbed run.

Additionally, we collected 161 web-based measurement results with at least ten repetitions per delay configuration.
Results cover nine browsers in 22 versions run on seven operating systems. %
That amounts to a total of 33 different combinations (see \Cref{tbl:browser_versions}).

\begin{table}
\begin{threeparttable}
    \small
    \caption{\ac{he} feature evaluation of client applications with our local and web-based measurement tools and their consistency between the measurement methods.}
    \label{tab:resultoverview}
    \begin{tabular}{lccccrrccp{0.75cm}}
        \toprule
                & \rlap{\rotatebox{60}{Prefers IPv6}} & \rlap{\rotatebox{60}{CAD Impl.}} & \rlap{\rotatebox{60}{\aaaar first\tnote{1}}} & \rlap{\rotatebox{60}{RD Impl.}} & \rlap{\rotatebox{60}{IPv4 Addrs. Used}} & \rlap{\rotatebox{60}{IPv6 Addrs. Used}}  & \rlap{\rotatebox{60}{Addr. Selection\tnote{2}}} &\rlap{\rotatebox{60}{Consistency}} &\\ %
        \midrule
        Chrome \emph{130.0}   & \tabbullet{} & \tabbullet{} & \tabbullet{} & \tabcirc{}\tnote{4}   &  1 &  1 &   \tabcirc{} &   \tabbullet{} &\\
        Chromium \emph{130.0}        & \tabbullet{} & \tabbullet{} & \tabbullet{} & \tabcirc{}\tnote{4}   &  1 &  1 &   \tabcirc{} &   \tabbullet{} &\\
        Edge \emph{130.0}            & \tabbullet{} & \tabbullet{} & \tabbullet{} & \tabcirc{}   &  1 &  1 &   \tabcirc{} &   \tabbullet{} &\\
        Firefox \emph{132.0}        & \tabbullet{} & \tabbullet{} & \tabcirc{} & \tabcirc{}   &  1 &  1 &   \tabcirc{} & \tableftcirc{} &\\
        Safari \emph{17.6}         & \tabbullet{} & \tabbullet{} & \tabbullet{} & \tabbullet{} & 10 & 10 & \tabbullet{} &     \tabcirc{} &\\
        \midrule
        Mobile Safari   & \tabbullet{} & \tabbullet{} & \tabbullet{} & \tabbullet{} &  - &  - &         - &           - &\\
        Chrome Mobile   & \tabbullet{} & \tabbullet{} & \tabbullet{} & \tabcirc{}   &  - &  - &         - &           - &\\
        \midrule
        \texttt{curl} \emph{7.88.1}           & \tabbullet{} & \tabbullet{} & \tabbullet{} & \tabcirc{}   &  1 &  1 &   \tabcirc{} &           - &\\
        \texttt{wget}\tnote{3} \emph{1.21.3}   & \tabbullet{} & \tabcirc{}   &   \tabcirc{} & \tabcirc{}   &  - &  1 &   \tabcirc{} &           - &\\
        \bottomrule
    \end{tabular}
    \begin{tablenotes}[flushleft]
        \footnotesize
        \item \tabbullet{} Observed as defined; \tableftcirc{} Observed with RFC deviation; \tabcirc{} Not observed
        \item[1] Can also depend on the operating systems stub resolver except for Chromium-based browsers which use their own stub resolver.
        \item[2] No web test validation. \tnote{3} Does not implement any type of \ac{he}.
        \item[4] Offers a flag to enable feature (\hevthree flag)
    \end{tablenotes}
\end{threeparttable}
\end{table}

\begin{figure}
    \includegraphics{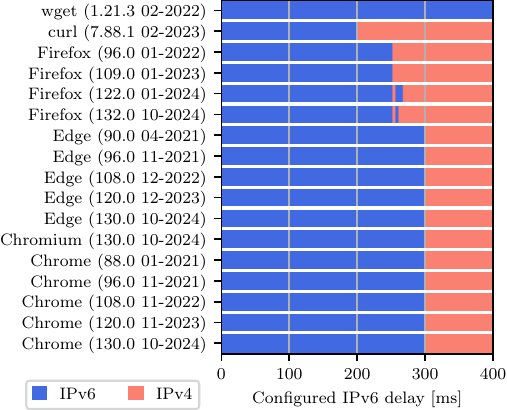}
    \caption{\reviewfix{1.1}\reviewfix{5} IP address family of established connection as measured on our local testbed setup. We omitted Safari, which has a \cad of \SI{2}{\second} to improve the visibility of the remaining results.}
    \label{fig:localcad}
\end{figure}

\Cref{tab:resultoverview} provides an overview of our results.
While all client applications prefer IPv6 if both versions are offered, \texttt{wget} (1.21.3) does not implement any fallback mechanism and thus does not implement \ac{he}.
\texttt{wget} fails without using the provided IPv4 addresses for delays larger than its timeout.
\Cref{fig:localcad} shows the address family of the established connection at each configured delay for all browsers and versions included in our local testbed measurement.
\reviewfix{1} The observed \cad obtained from the packet captures overlaps with the largest delay on an established connection via IPv6.
Only Firefox has a few outliers, but the median and standard deviation are within a \si{\milli\second} of the obtained value.

All Chromium-based browsers use a \cad of \SI{300}{\milli\second}.
\reviewfix{1} This value can also be found in its code~\cite{chrome-300ms-code}.
Measurement results collected via our web-based tool for these browsers (including mobile variants) validate these observations.
The command-line tool \texttt{curl} uses the smallest \cad of \SI{200}{\milli\second} (\reviewfix{1}see code~\cite{curl-code-he,curl-man-he}). %
Firefox follows the RFC recommendation of \SI{250}{\milli\second}, but has a few outliers where it waits longer than \SI{250}{\milli\second} for the IPv6 connection establishment.

In contrast to other browsers that matched our local measurements, we observed a wide range of \cad values for Safari~\cite{apple-he} in our web-based measurements.
While local measurements were consistent at \SI{2}{\second}, they ranged from \SI{50}{\milli\second} up to \SI{5}{\second} on the web, indicating a dynamic application of \cad.
Moreover, Safari also showed a unique behavior.
While other web-based measurements usually show a clear picture of IPv6 usage up until the \cad value, the measurement results with Safari did not.
This behavior manifested in repetitions where it used IPv4 at a smaller delay and IPv6 again for larger delays.
Firefox and Chrome also had a few test runs with such inconsistencies, but these only manifested in a maximum of two out of ten test repetitions.
At the same time, Safari exposed them between six and ten times.
To investigate this behavior we conducted web-based measurements with different network conditions.
Neither the network context (cable vs wireless, the access network, parallel network activity), nor the focus of the application window (MacOS and Safari can throttle background tabs and windows), nor the power supply did have any noticeable impact on the reproducibility of results.
Mobile Safari measurements report varying \cad values and inconsistency.
Nevertheless, the \cad never rose beyond \SI{1}{\second} for IPv6 connections on mobile phones with iOS.
We assume a preference towards lower values on mobile devices.
Our web-based measurements can help to evaluate these effects in more detail with different clients and conditions.

Moreover, Apple announced and deployed iCloud Private Relay (iCPR), a privacy-preserving proxying service, in December 2021~\cite{privaterelay, Sattler2022ICPR}.
As part of iCPR, Safari can relay traffic via an egress operator using MASQUE proxying~\cite{rfc9298, rfc9484}.
Connections via iCPR show completely different \ac{he} behavior compared to Safari.
The reason is that Safari does not create an IP tunnel via the relay network but only informs the egress node about the server name it wants to connect to.
The egress node handles the DNS resolution and the stack up to the transport protocol headers of TCP and UDP.
Therefore, measurements via iCPR show how the egress operators implement \ac{he}.
Akamai and Cloudflare egress nodes use a \cad of \SI{150}{\milli\second} and \SI{200}{\milli\second}, respectively.
\reviewfix{4} We cannot actively select the egress node provider and Fastly did not appear as a provider in our measurements.
This example shows that local testbed measurements can provide an overview of which features are deployed, but a web-based measurement campaign must be used to validate the results to cover complex real-world scenarios.

\subsection{Resolution Delay}
\label{sec:rd}

The \rd test case delays the \aaaar record to observe how clients implement the \rd{}.
Already, the testbed measurement results show that only Safari actually implements it.
We found Safari to be using the RFC recommendation of \SI{50}{\milli\second}.
In their default state, Chromium-based browsers and Firefox depend on the resolver's timeout.
They do not apply any DNS resolution timeout on their own.

The missing timeout leads to the delegation of timeouts to resolvers.
Only after the \aaaar query hits the resolver's timeout do these browsers initiate the IPv4 connection.
Similar observations are valid for \texttt{curl} and \texttt{wget}.

Usually, in the context of \ac{he}, IPv4 is considered the stable fallback, while IPv6 is the \emph{new} variant that can break.
In reality, IPv4 is not \sperc{100} reliable, and queries for \ar records are also sent via IPv6.
Therefore, we test how clients react to delayed \ar query responses.
To our astonishment, while all clients continued to prefer IPv6, all but Safari always waited for the \ar response to arrive.
Only then would they initiate the IPv6 connection to the target host, although the \aaaar record was not delayed.
This behavior resulted in Chrome and Firefox completely failing connections in case of high delays with some resolver configurations.
Since April 2024, Chromium provides a feature flag (\hevthree) that adds \rd{} and gets rid of this problematic behavior~\cite{chome-flag-initial,chrome-flag-hev3}.

Similar to the \cad evaluation, we collected web-based measurements with Safari via iCPR.
We observed that Cloudflare egress nodes use IPv6 up until a delay of \SI{1.75}{\second}.
Akamai egress nodes use a timeout of \SI{400}{\milli\second}.
Both operators use the same timeout for \ar and \aaaar record queries.
These properties are far from the behavior of Safari without iCPR.
Therefore, iCPR has the potential to negatively influence the user experience if issues on the resolution path exist.

The address selection is another measure standardized as part of \ac{he}.
We measured it with a domain name with 10 \ar and \aaaar records.
All records pointed to unresponsive addresses.
In \Cref{tab:resultoverview}, we list the number of addresses per IP family used for connection attempts by the clients.
Again, only Safari shows a visible address selection behavior (see App. \Cref{fig:he_attempted_addrs}).
Others only perform a simple fallback to IPv4 and stop after that, ignoring other records' addresses.

\subsection{Recursive Resolvers}
\label{sec:recursive_resolver}

Most browsers do not implement \rd and rely on the resolver's timeout.
To understand how and if resolvers implement some version of IPv6 version preference, we evaluate the behavior towards authoritative name servers.

\begin{table}
    \begin{threeparttable}
        \small
        \caption{Evaluation of resolver IPv6 usage as observed on the authoritative name server. For open resolvers, the observed IPv6 delay is a lower bound.}
        \label{tab:resolverstats}
        \begin{tabular}{lcrrr}
            \toprule
                & \aaaar & IPv6\ & Max. IPv6 & \# IPv6 \\
            Service  & Query & Share & Delay Used & Packets \\
            \midrule
            BIND & \tableftcirc & \sperc{100.0} & \SI{800}{\milli\second} & 1 \\
            Unbound & \tabbullet{} & \sperc{43.8} & \SI{376}{\milli\second} & 2 \\
            Knot Resolver & \tabdowncirc{} & \sperc{27.9} & \SI{400}{\milli\second} & 2 \\
            \midrule
            DNS.sb & \tableftcirc{} & \sperc{0.0} & - & - \\
            Google P. DNS & \tabrightcirc{} & \sperc{0.0} & - & - \\
            DNS0.EU & \tabbullet{} & \sperc{9.5} & -\tnote{1} & 2 \\
            NextDNS & \tabbullet{} & \sperc{8.9} & \SI{200}{\milli\second} & 1 \\
            Quad 101 & \tabbullet{} & \sperc{10.0} & \SI{400}{\milli\second} & 1 \\
            114DNS & \tabbullet{} & \sperc{11.1} & \SI{600}{\milli\second} & 1 \\
            Cloudflare & \tabbullet{} & \sperc{11.1} & \SI{500}{\milli\second} & 2 \\
            Verisign P. DNS & \tabbullet{} & \sperc{15.3} & \SI{250}{\milli\second} & 1 \\
            Yandex & \tabbullet{} & \sperc{17.4} & \SI{300}{\milli\second} & 6 \\
            H-MSK-IX & \tabbullet{} & \sperc{20.5} & \SI{600}{\milli\second} & 2 \\
            MSK-IX & \tabbullet{} & \sperc{22.1} & \SI{600}{\milli\second} & 2 \\
            Quad9 DNS & \tabbullet{} & \sperc{34.2} & \SI{1250}{\milli\second} & 2 \\
            OpenDNS & \tabbullet{} & \sperc{100.0} & \SI{50}{\milli\second} & 1 \\
            \bottomrule
        \end{tabular}
        \begin{tablenotes}[flushleft]
            \footnotesize
            \item \tabbullet{} Sends \aaaar before \ar{}; \tableftcirc{} Sends \aaaar after \ar{}; \tabrightcirc{} Sends \aaaar after query to IPv4 auth. name server; \tabdowncirc{} Sends either \aaaar or \ar but not both
            \item[1] Cannot determine the delay due to parallel queries on IPv4 and IPv6.
        \end{tablenotes}
    \end{threeparttable}
\end{table}

We focus on three popular recursive resolvers for our local measurements: BIND 9~\cite{bind}, Unbound~\cite{unbound}, and Knot resolver~\cite{knotresolver}.
Furthermore, we issued queries to 82 IP addresses of 17 popular open resolvers (see App. \Cref{tab:resolvers}). %
While our sample does not cover ISP-operated resolvers, it shows the differences that resolvers can expose. %
Moreover, we provide a web-based testing tool that allows users to check their configured resolver.
Our analysis only includes resolvers that can resolve zones with IPv6-only authoritative name servers.
Hurricane Electric, Lumen (Level3), Dyn, and G-Core are not able to resolve domain names with an IPv6-only delegation, leaving us with 13 to analyze.
For Quad 101, only the IPv6 addresses successfully connect to our IPv6-only name servers.
One \texttt{dns0.eu} IP address did not provide reliable IPv6-only resolution.

To connect to our IPv6 addresses, the resolvers need to obtain the address through glue records or name resolution.
All operators sent \aaaar queries for the name server names. %
Even when we include glue records, 12 out of 13 public services and the BIND resolver requested the \aaaar record before issuing the query to the authoritative name server.
Only Google Public DNS does not issue any \aaaar query before sending the query to the authoritative name server.
The Knot resolver does send out either \ar or \aaaar queries for the delegated name server name but never both.

11 open resolvers and Unbound sent the \aaaar record before the \ar record, a behavior defined in RFC 8305~\cite{rfc8305}.
DNS.sb, Google Public DNS, and the BIND resolver query the \ar record first.
While packet reordering could influence our results, the results were consistent over different test runs and delay configurations.

When the resolver has both addresses available, it needs to select an address for the next iterative query.
BIND performs classic \ac{he} IP version preference and always prefers IPv6.
Knot and Unbound use IPv6 in \sperc{25} and \sperc{50,} respectively (see \Cref{tab:resolverstats}).
We observe a \cad of \SI{800}{\milli\second} for BIND, \SI{400}{\milli\second} for Knot, and \SI{376}{\milli\second} for Unbound.
While BIND and Knot are consistently falling back to IPv4 addresses, Unbound retries the IPv6 address in \sperc{44} of the times and increases the \cad to \SI{1128}{\milli\second} due to an exponential backoff~\cite{unboundtimeout}.

Except for Google Public DNS and DNS.sb, all public operators use the IPv6 name server address at least once.
Only OpenDNS always tries to connect via IPv6 first and falls back to IPv4 after \SI{50}{\milli\second}.
The other operators do not perform any type of \ac{he} style behavior.
Nine operators do not use IP protocol version interleaving and send up to six queries to the IPv6 address (Yandex).
We never see DNS0.EU switching IP protocol version when retrying, but it sticks to the IP version initially chosen and fails at some point.

\section{Discussion and Conclusion}
\label{sec:conclusion}

We systematically approach the measurement of the client's \acl{he} feature implementations.
A local testbed allows us to perform in-depth measurements in a controlled setup.
The web-based measurements can validate local results and cover more clients in actual network conditions.
All widely used browsers prefer IPv6 connections when possible.
Except for Safari, which employs a dynamic, unpredictable approach, the observed \cad values are within a \SI{50}{\milli\second} interval of the \SI{250}{\milli\second} RFC suggestion.

Only Safari implements a \rd and address selection mechanisms, thus, \ac{he}v2.
In contrast, Chromium-based browsers and Firefox delay IPv6 connection initialization until the corresponding DNS \ar record arrives, even though IPv6 address information is available and preferred.
We suggest implementing a timeout for DNS queries for all clients, even when \ac{he} is not implemented.
The current situation is even worse from an IPv6 deployment perspective, as slow \ar queries also slow down IPv6, even if it is not at fault.

Moreover, not only the browser software itself but also new proxying techniques can impact the \ac{he} behavior.
This result shows the usefulness of our web-based measurement tool in understanding the ecosystem in different scenarios, how clients with different configurations currently behave, and how future updates (\eg the Chrome feature flag) change behavior.
For example, we determined that if iCPR is enabled, Safari users lose \rd and address selection features through our web test.

Due to the reliance on resolvers and their significant role, we included them in our analysis.
Only single instances and applications follow an HE-style approach.
We suggest starting dedicated discussions to develop recommendations on the behavior of protocol preference for critical Internet infrastructure clients, such as DNS resolvers.
You can find our web-based tool and our measurement tools' source code on: \url{https://www.happy-eyeballs.net/}.

\begin{acks}
    We thank the anonymous reviewers and our shepherd Pawel Foremski for their valuable feedback.
    This work was partially funded by the German Federal Ministry of Education and Research under project PRIMEnet (16KIS1370).
\end{acks}

\label{body}

\bibliographystyle{ACM-Reference-Format}
\balance
\bibliography{refs}

\appendix

\begin{figure*}
    \includegraphics{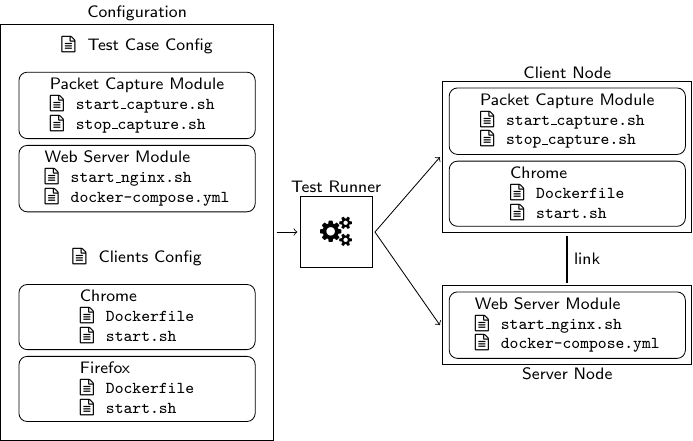}
    \caption{Overview of the local testbed and its components and configurations (see \Cref{sec:approach}).}
    \label{fig:local_testbed}
\end{figure*}
\section{Ethics}
\label{sec:ethcis}

We follow strict ethical measures in our recursive resolver active measurements and our web-based measurement campaign.
These measures are mainly based on informed consent~\cite{menloreport} and well known best practices~\cite{PA16}.
We apply measures described by \citeauthor{durumeric_zmap_2013} \cite{durumeric_zmap_2013} to our active recursive resolver measurements.
We limit the rate of our scans and use a blocklist based on requests to be excluded from our scans.
The scanning prefix IRR information points to an email contact on our control to react quickly to all requests.
Furthermore, scanning machines operate a website pointing to information about our scans.
We did not receive any complains about this research.

The web-based measurements in this paper were only collected by research group members and devices owned by them.
We collect information on the client and its state during the web clients measurement to get a better picture.
This includes the client's user agent and plattform as provided by the browser and the client network's \ac{as} number which are not considered personal identifiable information.
We need this information to correctly attribute results to applications, OSes, and access networks.
The findings for Safari with iCloud Private Relay were only possible by including the \ac{as} information.
Nevertheless, we offer users the option to not submit results to our study and hence no data is collected while using the tool.

\section{Framework Details}

\Cref{fig:local_testbed} provides an overview about our \textit{Local Testbed Framework} as described in \Cref{sec:approach}.
Test cases, the host setup and tested clients are specified and adapted in one configuration.
A test runner sets up the client and server node accordingly and iterates through all test cases, configurations and clients.

\begin{figure}
    \begin{subfigure}{\linewidth}
        \includegraphics[width=\linewidth]{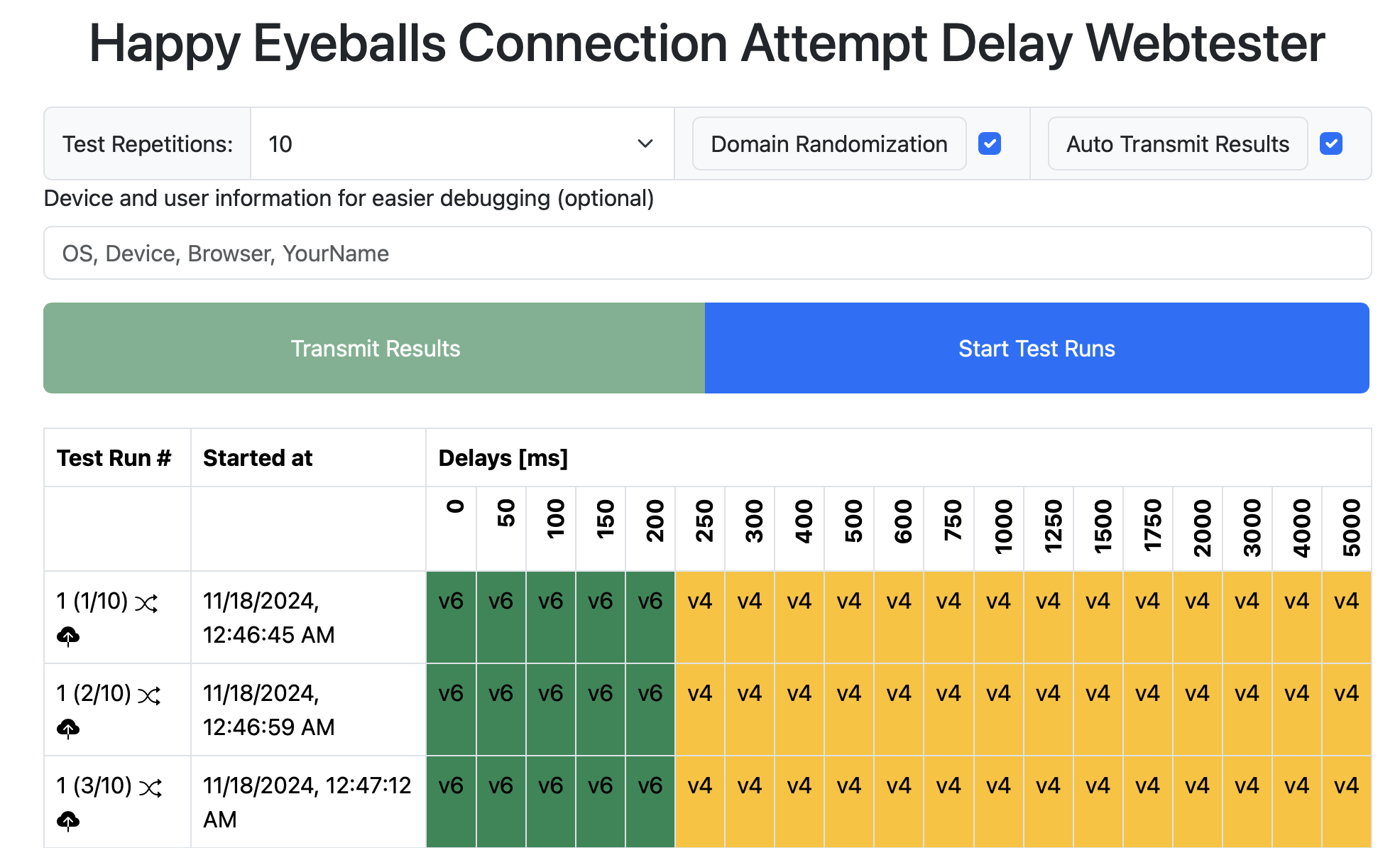}
        \caption{\acl{cad} testing tool website.}
        \label{fig:cadweb_screenshot}
    \end{subfigure}%

    \begin{subfigure}{\linewidth}
        \includegraphics[width=\linewidth]{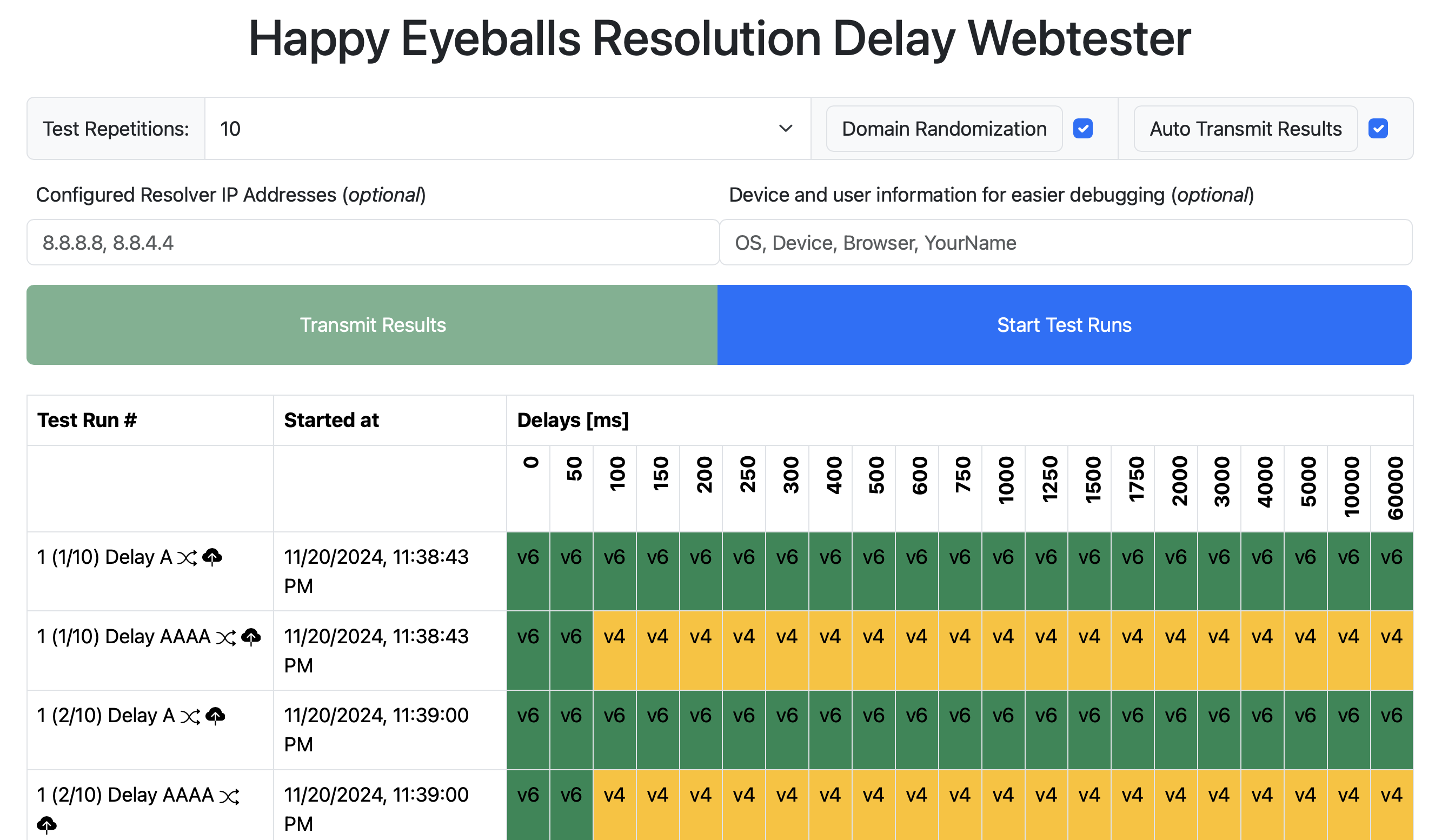}
        \caption{\acl{rd} testing tool website.}
        \label{fig:rdweb_screenshot}
    \end{subfigure}
    \caption{Screenshots of our web-based testing tool website taken from measurements with Safari (see \Cref{sec:approach}). We make the tool publicly available on accept to allow individuals to easily test and understand their client behavior.}
    \label{fig:web_screenshots}
\end{figure}

\Cref{fig:web_screenshots} shows an example result of our \cad and \rd web-based testing tool as described in \Cref{sec:approach}.
It allows users testing their individual setup (browser + resolver + OS) and their \ac{he} capabilities.
The source code for the test framework and the web-based tool are publicly available.
Both tools include a complete Ansible setup script.
We will make the tool publicly accessible if the paper is accepted.

\section{Open Resolvers}
We tested the \ac{he} capabilities of open resolvers listed in \Cref{tab:resolvers}.
We selected these from a publicly available list of open resolvers\footnotemark.
Results regarding the capabilities of these resolvers are described in \Cref{sec:recursive_resolver}.

\footnotetext{\url{https://gist.github.com/mutin-sa/5dcbd35ee436eb629db7872581093bc5}}

\newpage
114DNS does only offer IPv4 resolver addresses but is capable of resolving domain names on IPv6-only resolution paths.
The WhoAmI tool by Akamai (\url{whoami.ds.akahelp.net}) shows an address in a different \ac{as}.
Therefore, we assume it is a forwarder and uses an IPv6-capable recursive resolver.

\begin{table}
    \caption{Tested recursive resolvers. Hurricane Electric, Lumen (Level3), Dyn, and G-Core do not support IPv6 on the resolution path and are therefore not included in our evaluation presented in \Cref{sec:recursive_resolver}.}
    \label{tab:resolvers}
    \small
    \centering
    \begin{tabular}{lrr}
        \toprule
        & \multicolumn{2}{c}{\# Addrs.} \\
        \cmidrule(lr){2-3}
        Service & IPv4 & IPv6 \\
        \midrule
        DNS.sb              & 2 & 2 \\
        Google P. DNS       & 2 & 2 \\
        DNS0.EU             & 2 & 2 \\
        114DNS              & 2 & 0 \\
        Cloudflare          & 2 & 2 \\
        H-MSK-IX            & 2 & 2 \\
        MSK-IX              & 2 & 2 \\
        NextDNS             & 2 & 2 \\
        OpenDNS             & 6 & 6 \\
        Quad 101            & 2 & 2 \\
        Quad9 DNS           & 6 & 6 \\
        Verisign P. DNS     & 2 & 2 \\
        Yandex              & 2 & 2 \\
        \midrule
        G-Core              & 2 & 2 \\
        DYN                 & 2 & 0 \\
        Lumen (Level3)      & 4 & 0 \\
        HE                  & 4 & 4 \\
        \bottomrule
    \end{tabular}
\end{table}

\section{Address Selection}
\begin{figure}
    \includegraphics[width=\linewidth]{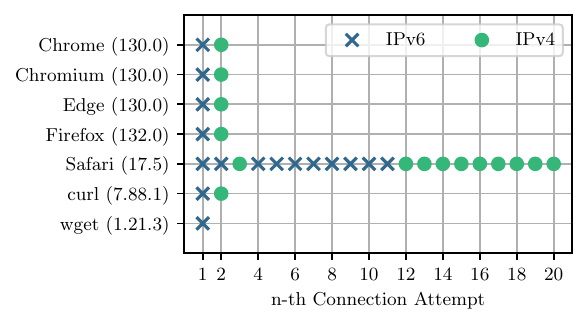}
    \caption{Address selection based on an internal priority list. The plot shows the address family used by browsers at the n-th connection attempt.}
    \label{fig:he_attempted_addrs}
\end{figure}

The main findings regarding address selection are described in \Cref{sec:rd}.
The following provides further information on the actually visible selection from different applications.
\Cref{fig:he_attempted_addrs} plots the address family used by browsers on parallel connection attempts when the browser hits its configured \cad.
We configured a total of ten addresses per family which do not offer a service but all time out.
Only Safari retries as often as there are addresses.
All others which implement the \cad only try one IPv6 and one IPv4 address.
This is the described behavior in \hevone{}.

Safari does use a \emph{First Address Family Count} of two and prefers IPv6.
Safari's interleaving strategy is to attempt one IPv4 address after the two IPv6 addresses.
Then it continues with all remaining IPv6 addresses and only after the remaining IPv4 addresses are tried.

\section{Measured Browser and OS Versions}

\begin{table}
    \caption{Operating systems and browsers for which we obtained web-based measurements. This information was extracted from the user agent. Therefore, some entries do not contain an OS version.}
    \label{tbl:browser_versions}
    \centering
    \begin{tabular}{lrlr}
        \toprule
        \multicolumn{2}{c}{OS} & \multicolumn{2}{c}{Browser} \\
        \cmidrule(lr){1-2}\cmidrule(lr){3-4}
        Name & Version & Browser & Browser Version \\
        \midrule
        Android & 10 & Chrome Mobile & 127.0.0 \\
        Android & 10 & Chrome Mobile & 130.0.0 \\
        Android & 10 & Firefox Mobile & 131.0 \\
        Android & 10 & Samsung Internet & 26.0 \\
        Android & 14 & Firefox Mobile & 125.0 \\
        Android & 14 & Firefox Mobile & 128.0 \\
        Android & 14 & Firefox Mobile & 131.0 \\
        Chrome OS & 14541.0.0 & Chrome & 129.0.0 \\
        Linux &  & Chrome & 130.0.0 \\
        Linux &  & Firefox & 128.0 \\
        Linux &  & Firefox & 130.0 \\
        Linux &  & Firefox & 131.0 \\
        Linux &  & Firefox & 132.0 \\
        Mac OS X & 10.15 & Firefox & 128.0 \\
        Mac OS X & 10.15 & Firefox & 131.0 \\
        Mac OS X & 10.15 & Firefox & 132.0 \\
        Mac OS X & 10.15.7 & Chrome & 127.0.0 \\
        Mac OS X & 10.15.7 & Chrome & 129.0.0 \\
        Mac OS X & 10.15.7 & Chrome & 130.0.0 \\
        Mac OS X & 10.15.7 & Opera & 114.0.0 \\
        Mac OS X & 10.15.7 & Safari & 17.4.1 \\
        Mac OS X & 10.15.7 & Safari & 17.5 \\
        Mac OS X & 10.15.7 & Safari & 17.6 \\
        Mac OS X & 10.15.7 & Safari & 18.0.1 \\
        Ubuntu &  & Firefox & 128.0 \\
        Ubuntu &  & Firefox & 131.0 \\
        Windows & 10 & Chrome & 127.0.0 \\
        Windows & 10 & Edge & 130.0.0 \\
        Windows & 10 & Firefox & 130.0 \\
        iOS & 17.5.1 & Mobile Safari & 17.5 \\
        iOS & 17.6 & Mobile Safari & 17.6 \\
        iOS & 17.6.1 & Mobile Safari & 17.6 \\
        iOS & 18.1 & Mobile Safari & 18.1 \\
        \bottomrule
        \end{tabular}
\end{table}

\reviewfix{3-1}
\Cref{tbl:browser_versions} contains the browser versions and the operating systems running these browsers used in our web-based measurements.
We extract this information from the browser provided user agent string.
Linux and Ubuntu do not provide the OS version in the user agent.

\label{lastpage}

\end{document}